\documentclass[twocolumn,twoside,slac]{revtex4}
\usepackage{graphicx}
\usepackage{fancyhdr}
\def\vev#1{\langle#1\rangle}

\begin{document}

\title{Statistical Issues in Particle Physics -- A View from BaBar}

%

\author{F. C. Porter}
\affiliation{(For\hbox{ the BaBar collaboration)}\\
 Lauritsen Laboratory of Physics, Caltech, Pasadena, CA 91125, USA}

\begin{abstract}
The statistical methods used in deriving physics results in the BaBar collaboration are reviewed, with especial emphasis on areas where practice is not uniform in particle physics. 
\end{abstract}

\maketitle

\smash{\hskip15cm\hbox{\raise4.5cm\vbox{\hbox{CALT 68-2463}\hbox{SLAC-PUB-10243}}}}

\thispagestyle{fancy}

\section{Introduction}

The purpose of the BaBar experiment at the PEP-II accelerator at SLAC is to study $e^+e^-$ collisions in the 10 GeV center-of-mass region, namely the region around $B\bar B$ threshold. In particular the program is to investigate extensively $CP$ violation and rare decays of $B$ mesons, as well as topics in charm and tau physics.

Here, BaBar's approach to statistical issues is summarized. Emphasis is given to areas which are often controversial.

\section{BaBar Analysis Organization}

\begin{figure}
\includegraphics[width=65mm]{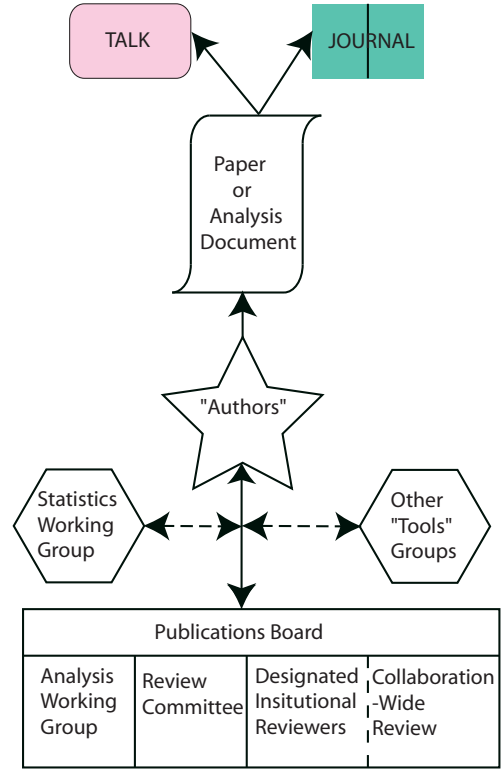}
\caption{BaBar analysis organization. A detailed analysis for some physics result is typically performed by a subset of the collaboration, labelled ``authors'' here. There are several layers of review that occur as an analysis moves towards publication: an Analysis Working Group interacts with the authors from the earliest stages; once a document is produced, a Review Committee of typically three people is assigned by the Publications Board to critically examine the analysis; upon approval from the Review Committee, the paper is circulated for collaboration-wide review, including several institutions specially designated to look closely at it. Oversight of the process and final review is carried out by the Publications Board.}
\label{fig:physorg}
\end{figure}

BaBar is a collaboration of approximately 600 physicists, from $\sim80$ institutions in a dozen countries. Thus, managing the production of physics results, from initial analysis to final publication, while maintaining collaboration involvement is a daunting task. An organizational structure has been established to facilitate this process, as illustrated in Fig.~\ref{fig:physorg}.

The ``Statistics Working Group'' was appointed by the Publications Board in order to provide guidelines and advice on statistical matters~\cite{bib:SWiG}. 
This group is advisory; I'll note how well the guidelines are actually adopted in some cases.

\section{Philosophy}
\label{sec:philosophy}

The approach to choosing a statistical procedure is to start by considering the goal. We adopt the view that there are two broad domains in terms of goal:

\begin{itemize}
 \item The first goal is that of summarizing the relevant information in a measurement. This is ``descriptive'' statistics. It is considered obligatory to report such a description of the result of the experiment. Inherent in this is the view that it is actually useful to do so, a notion that is not uniformly accepted. The use of frequency statistics is recommended for this purpose. The choice within the domain of possible frequency statistics is driven by an emphasis on clarity and the facility to compare and combine with other measurements.
 \item The second goal is that of interpreting the relevant information in the context of making a statement about ``physics''. This is regarded as optional, since once the relevant information is available people are in principle able to do this step for themselves. Because a statement about physical reality may depend on other information, and on theoretical input, Bayesian statistics are recommended.  
\end{itemize}

It may be remarked that there may be other goals, such as making a decision concerning how to spend money for the next experiment. This would involve, beyond the above interpretive aspects, a consideration of the risks and benefits. We take the point of view that this is outside the scope of the analysis and reporting of results, and hence do not discuss it further. 

\section{Statistical Practice in BaBar}

We turn now to a review of the specific statistical practices recommended or adopted in BaBar analyses. Not included are the methods and tools used for optimizing analyses, and pattern recognition, data reduction, and simulation procedures. These matters are crucial, but here we emphasize instead areas which are traditionally more controversial. It should be mentioned that the typical products of a BaBar physics analysis are:
\begin{enumerate}
 \item ``Best'' estimates for physical parameters.
 \item Interval estimates for physical parameters.
 \item Significance levels of observations (e.g., of a possible discovery).
 \item Goodness-of-fit of models to the data.
\end{enumerate}

\subsection{Blind Analysis}

Many BaBar results are obtained in ``blind analyses''. The purpose of a blind analysis is to avoid the introduction of bias, which could occur if the analyst is looking at the results as the analysis is designed. There is more than one approach to ``blindness'', see the talk by Aaron Roodman~\cite{bib:Roodman} for a summary of BaBar practice. We'll give one example here.

\begin{figure}
\includegraphics[width=78mm]{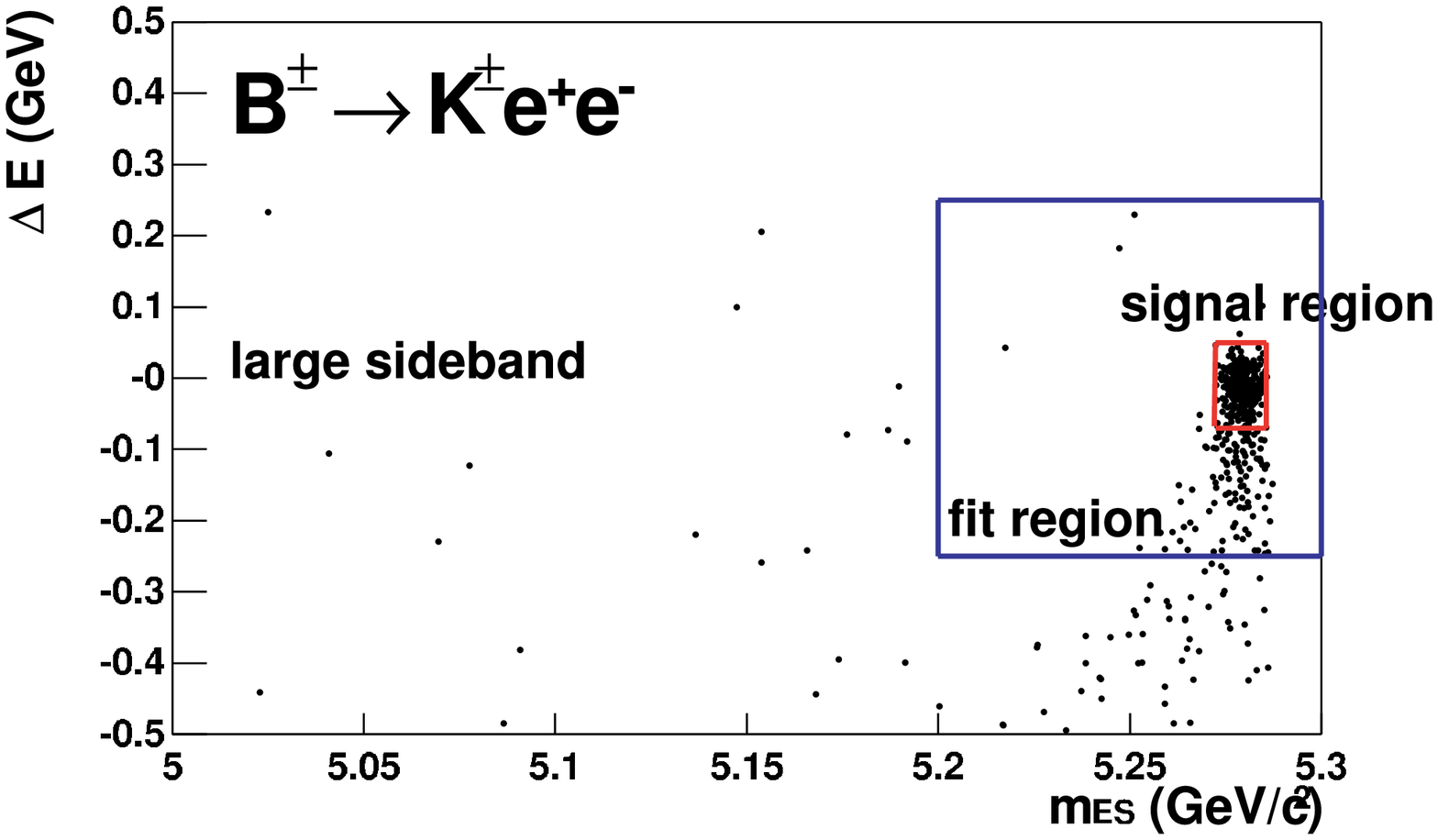}
\includegraphics[width=75mm]{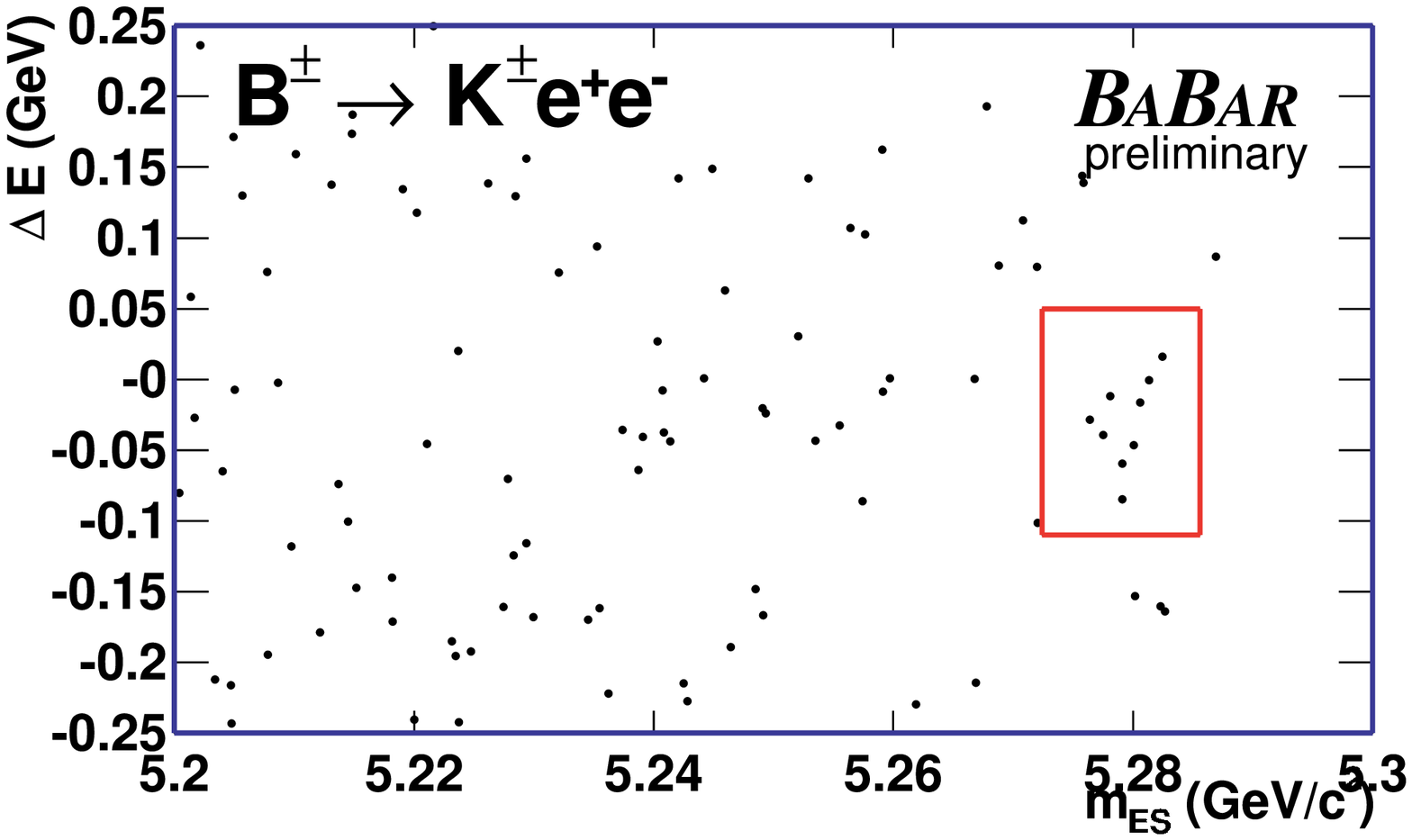}
\caption{Example of a blind analysis in BaBar. The upper plot shows a Monte Carlo simulation of the signal $B^\pm\to K^\pm e^+e^-$ process. The outside boundaries delimit the ``large sideband region''; the intermediate box is the ``fit region'', and the inner box is the region in which the signal is concentrated (referred to as the ``signal region'', but in fact playing no special role in the analysis). The lower plot shows the BaBar data after unblinding. Here, the outside boundaries demarcate the fit region, and the smaller box is the ``signal region''.}
\label{fig:kee}
\end{figure}

For example, consider the measurement of the rare $B$ decay $B^\pm\to K^\pm e^+e^-$~\cite{bib:Walsh}, of interest because of its sensitivity to possible physics beyond the standard model. The basic idea of the analysis is to look for a signal which peaks in the distribution of two kinematic variables, known as ``$\Delta E$'' and ``$m_{\rm ES}$'' (Fig.~\ref{fig:kee}). A fit is performed to this two-dimensional distribution in order to extract the strength of any signal present. However, before performing the fit, an event selection is made in order to suppress backgrounds. In order to avoid biasing the result by looking at the data while tuning the selection, a blind analysis is performed. 

The $\Delta E-m_{\rm ES}$ plane is divided into two regions: a region where the fit will be performed, which includes the region where a signal might appear; and a larger (``large sideband'') region which excludes the fit region. During the tuning of the analysis, the data may not be looked at in the fit region, only in the large sideband region. Monte Carlo and control sample data (including a type of data resembling signal) are used to tune the analysis. Once the selection criteria have been established, the fit region of the data is revealed, and the fit performed to extract the result.

As BaBar is continuing to accumulate data, an issue arises when it is desired to update a blind analysis to include new data. In principle, one could simply add the new data, without changing the analysis. However, this may be impractical, or undesirable. For example, the entire dataset may be re-reconstructed with improved constants or pattern recognition code. Or, there may have been improvements in tools such as particle identification. One would like to incorporate the benefit from such improvements. Additionally, it might be desirable to work harder to optimize the analysis, or to optimize on different criteria, such as precision instead of sensitivity. BaBar often takes a practical compromise approach to incoporate new data, and such improvements. We have the notion of ``re-blinding'' the data, and re-optimizing. It is considered safe in this re-optimization to use variables which have not been inspected too carefully in the blind region in the first dataset. Nonetheless, once we have done this, we do not refer to the new result as having been done with a blind analysis.

BaBar is perhaps the first large HEP collaboration to have embraced the blind methodology so enthusiastically. However, not every BaBar analysis is blind. In particular, analyses which may be called exploratory are generally not blinded. A recent example from BaBar is the discovery of the $D_{sJ}^*(2317)^\pm$~\cite{bib:Ds}, which was not the result of a blind analysis. There are many examples of people being led astray by such non-blind exploratory analyses, so extreme caution is warranted. The exploratory nature of such analyses makes it difficult to apply rigorous methodologies with well-defined statistical properties. It may not be impossible to do better though \cite{bib:Knuteson}.

\subsection{Confidence Intervals}

The recommendation in BaBar is to use frequency statistics for summarizing information (Sect.~\ref{sec:philosophy}). The goal is to describe what is observed, stressing simplicity and coherence of interpretation, as well as facility in combining with other results. With these criteria, we think it can be counter-productive to impose ``physical'' constraints. There is no reason to obscure the observation of an ``unlikely'' result. Imposing constraints may also complicate combination of results. Generally, the recommendation is to quote two-sided 68\% confidence intervals as the primary result. Where there may be doubt, a check for frequency validity (coverage) should be performed.

\subsubsection{Example in Two Dimensions}
\label{sec:Dmix}

As an example of the construction of a confidence region in a BaBar analysis, consider the measurement of $D$ mixing and doubly Cabibbo suppressed $D$ decays~\cite{bib:Dmix}. In this analysis, two parameters of interest are to be determined, which may be expressed as $x^\prime$ and $y^\prime$ according to the relations:
\begin{eqnarray}
 x^\prime &\equiv & {\Delta m\over \Gamma}\cos\delta +{\Delta\Gamma\over 2\Gamma}\sin\delta, \\
 y^\prime &\equiv & {\Delta \Gamma\over 2\Gamma}\cos\delta -{\Delta m\over \Gamma}\sin\delta,  
\end{eqnarray}
where $m$ and $\Gamma$ are the $D$ mass and width, $\Delta m$ and $\Delta \Gamma$ are the (small) differences in masses and widths between the two $D$ mass eigenstates, and $\delta$ is an unknown strong phase (between Cabibbo-favored and doubly Cabibbo suppressed amplitudes). The measurement is only sensitive to $x^{\prime 2}$ and $y$, and it is possible that the maximum of the likelihood will occur at $x^{\prime 2}<0$ (``unphysical'' region). At the current level of sensitivity, we should find a result consistent with $x^{\prime 2}=y^\prime=0$, if the standard model is correct.

\begin{figure}
\includegraphics[width=80mm]{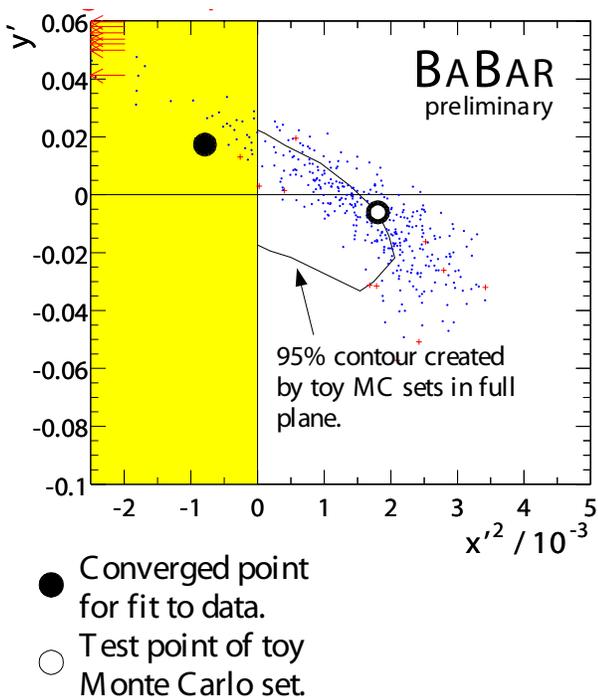}
\caption{Finding a confidence contour in two dimensions~\cite{bib:Egede}. The large filled dot shows the location of the maximum likelihood for the BaBar data. The open dot shows the value of $(x_0^{\prime 2}, y^\prime_0)$ chosen for a simulation. The small dots show simulated experiments for which
$\lambda_{\rm MC} > \lambda_{\rm Data}$. The pluses, as well as the arrows pointing offscale, show simulated experiments for which $\lambda_{\rm MC} < \lambda_{\rm Data}$. The 95\% contour resulting from the algorithm described in the text is shown. The shaded region is the ``unphysical'' region. Note that the evaluation of the maximum likelihood is not restricted to the ``physical'' region.}
\label{fig:dmixStat}
\end{figure}

The construction of a confidence region in the two-dimensional $(x^{\prime 2},y^\prime)$ plane, corresponding to 95\% confidence level with the frequency interpretation, is performed as follows (Fig.~\ref{fig:dmixStat}): 
\begin{enumerate}
\item Pick a point $(x_0^{\prime 2}, y^\prime_0)$ in the plane.
\item Form the ``data'' likelihood ratio comparing the observed maximum likelihood with the likelihood at $(x_0^{\prime 2}, y^\prime_0)$:
 \begin{equation}
  \lambda_{\rm Data} = {{\cal L}_{\rm max}({\rm Data}) \over {\cal L}_{(x_0^{\prime 2}, y^\prime_0)}({\rm Data})}.
 \end{equation}
\item Simulate many experiments with $(x_0^{\prime 2}, y^\prime_0)$ taken as the true values of the parameters.
\item For each Monte Carlo simulation form the ``MC'' likelihood ratio:
 \begin{equation}
  \lambda_{\rm MC} = {{\cal L}_{\rm max}({\rm MC}) \over {\cal L}_{(x_0^{\prime 2}, y^\prime_0)}({\rm MC})}.
 \end{equation}
\item From the ensemble of simulations, determine the probability $P(\lambda_{\rm MC} > \lambda_{\rm Data})$. If this probability is greater than 0.95, then the point $(x_0^{\prime 2}, y^\prime_0)$ is inside the contour; if less than 0.95, then the point is outside the contour.
\item This procedure is repeated for many choices of $(x_0^{\prime 2}, y^\prime_0)$ in order to map out the contour.
\end{enumerate}
Fig.~\ref{fig:dmixStat} shows the result of this algorithm. The choice was made to stop computng the contour at the border of the ``physical'' region. The computation could in principle have been carried into the ``unphysical'' region (up to technical difficulties of the sort we shall discuss anon). It of course makes no difference to the frequency interpretation whether it is extended into the ``unphysical'' region or not.

\subsubsection{Low Statistics Issues}

Issues arise in applying the recommendation of always quoting a two-sided interval for a parameter when the sampling is not from an approximate normal distribution. Most often this involves the low-statistics regime of a counting process.

\begin{figure}
\includegraphics[width=65mm]{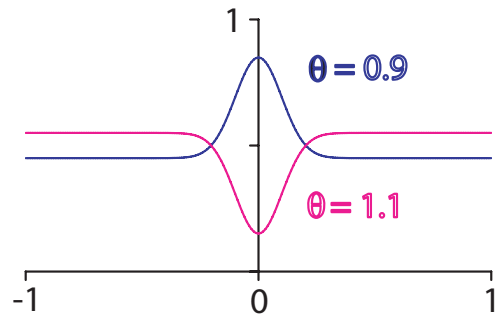}
\caption{Graph of the example sampling PDF for two values of parameter $\theta$: $\theta=0.9$, and ``unphysical'' (negative signal) value $\theta=1.1$. Note that both values are mathematically permissible.}
\label{fig:boundaryPDF}
\end{figure}

The first issue is a technical one: it can happen that a search in parameter space wants to go into a region where the probability distribution is undefined. This is distinct from going into an ``unphysical'' region as in the example above: we'll call it crossing a ``math boundary''. As a simple example, consider the case of a normal ``signal'' on a flat ``background'', with PDF (Fig.~\ref{fig:boundaryPDF}):
\begin{equation}
 p(x;\theta) = {\theta\over 2} + {1-\theta \over A\sqrt{2\pi}\sigma}e^{-{x^2\over 2\sigma^2}},\ x\in(-1,1).
\end{equation}
The parameter of interest is the strength of the signal, here expressed as $1-\theta$, the probability of sampling a signal event. An experiment samples $N$ events from this distribution, with likelihood function:
\begin{equation}
{\cal L}\left(\theta;\{x_i,i=1\ldots,N\}\right) = \prod_{i=1}^N p(x_i;\theta).
\end{equation}

It is quite possible that the likelihood will be maximal for a value of $\theta$ for which the PDF is not defined. The function $p(x;\theta)$ may become negative in some region of $x$. If there are no events in this region, the likelihood is still ``well-behaved''. 
However, the resulting fit, as a description of the data, will typically look poor even where the PDF is positive. This is considered unacceptable.

\begin{figure}
\includegraphics[width=75mm]{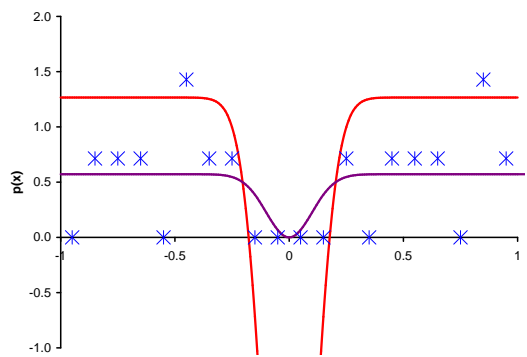}
\vskip-1cm\caption{Example of a possible dataset generated according to the flat background plus normal signal PDF. The data are displayed in histogram form by the points. The curve that goes negative (and is cut off at the plot boundary) is the result of the (unbinned) maximum likelihood fit. The other curve is the result of the same fit, except with the constraint that it cannot become negative.}
\label{fig:unMathRegion}
\end{figure}

An illustration of a possible sampled dataset from this distribution is shown in Fig.~\ref{fig:unMathRegion}, displayed as a histogram. An (unbinned) maximum likelihood fit to this data gives an estimate for $\theta$ in a region outside the math boundary. The graph of the ``PDF'' curve for this estimate does not give a good representation of the data. On the other hand, if the fit is constrained to the math region, the graph of the PDF curve looks like a reasonable representation of the data.

Thus, we suggest as a practical resolution to this problem to constrain the fit to remain within bounds such that the PDF is everywhere legitimate (n.b., parameters may still be ``unphysical'').
Experience is that this gives fits which ``look'' like the data, as in the present example, Fig.~\ref{fig:unMathRegion}.
This same practical recommendation applies in interval evaluation (but coverage should be checked, as always).

Another issue that arises frequently in low statistics (Poisson) sampling may be expressed in the form of the following example:
A ``cut and count'' analysis for a branching fraction $B$ finds $n$ events. The mean expected background contribution is estimated as $\hat b\pm\sigma_b$ events. The efficiency and parent sample are estimated to give a scale factor (relating observed signal events to $B$) of
$\hat f\pm\sigma_f$. The problem is to determine a confidence interval (at 68\% confidence, say), in the frequency sense, for $B$.

We'll assume that $n$ is sampled from a Poisson distribution with mean $\mu=\langle n \rangle = fB+b$, that $\hat b$ is sampled from a normal distribution, $N(b,\sigma_b)$, and that $\hat f$ is sampled from a normal distribution, $N(f,\sigma_f)$. Thus the likelihood function is:
\begin{equation}
{\cal L}(n, \hat b, \hat f; B, b, f) = {\mu^n e^{-\mu} \over n!} {1\over 2\pi\sigma_b\sigma_f}e^{-{1\over 2}\left({\hat b - b\over\sigma_b}\right)^2-{1\over 2}\left({\hat f - f\over\sigma_f}\right)^2}.
\end{equation}
It should be noted that this example is realistic, arising in practice (to a good approximation). A variant is to assume a normal distribution in $\widehat{(1/f)}$

Several methods have been proposed, and used, for dealing with this problem (see Ref.~\cite{bib:Barlow} for further discussion of these):
\begin{enumerate}
 \item Just give $n$, $\hat b\pm\sigma_b$, $\hat f\pm\sigma_f$. This provides a complete summary of the relevant information, and should be done anyway. But it isn't a confidence interval for $B$.
 \item Integrate out the nuisance parameters according to
\begin{eqnarray}
\hskip0.85cm{\cal L}(n, \hat b, \hat f; B) =&&\\
 &&\hskip-2.5cm \int df \int db {\mu^n e^{-\mu} \over n!} {1\over 2\pi\sigma_b\sigma_f}e^{-{1\over 2}\big({\hat b - b\over\sigma_b}\big)^2-{1\over 2}\big({\hat f - f\over\sigma_f}\big)^2}.\nonumber
\end{eqnarray}
This is easy, and often done. It may be interpreted as a partially Bayesian approach, where a uniform prior has been assumed for $f$ and $b$. The frequency properties could be investigated, but usually aren't.
 \item A very common approach when quoting upper limits is to do the appropriate Possion statistical analysis for $n$, but with the scale and background parameters fixed at the estimated values shifted by one standard deviation (in the direction to make the limit higher than with the central values). This has the benefit of being very easy to do, but it is clearly ad hoc, and the coverage is usually not investigated. 
\end{enumerate}
Here, I would like to comment on the possibility of evaluating these confidence intervals in another way.

The method I consider is actually a very common method that seems to have been rather neglected as an approach to the present problem. The algorithm is as follows: First, find the global maximum of the likelihood function with respect to $B, f,b$. Then search in the $B$ parameter for the point where $-\ln{\cal L}$ increases from the minimum by a specified amount (perhaps by $\Delta=1/2$ for a 68\% confidence interval), making sure that the likelihood is re-maximized with respect to $f$ and $b$ during this search. The resulting points $B_\ell,B_u$ then give an estimated interval for parameter $B$ which we would like to be a confidence interval.

\begin{figure}
\includegraphics[width=60mm,angle=270]{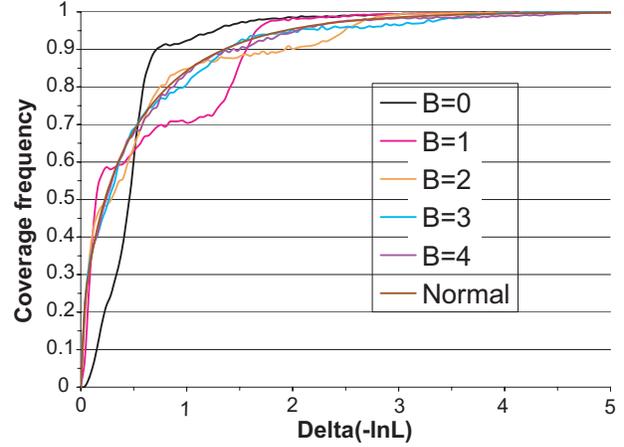}
\caption{Coverage frequency as a function of $\Delta$ for $f=1$, $\sigma_f=0.1$, $b=0.5$, $\sigma_b=0.1$. There are several curves corresponding to different numbers of expected signal events, $B$. The smoothest curve is the coverage in the high statistics (normal) limit.}
\label{fig:NM1}
\end{figure}

\begin{figure}
\includegraphics[width=60mm, angle=270]{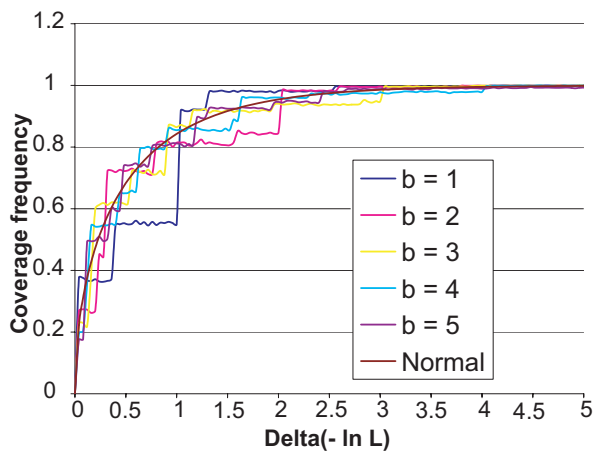}
\caption{Coverage frequency as a function of $\Delta$ for $B=0$, $f=1$, $\sigma_f=0$, $\sigma_b=0.1$. There are several curves corresponding to different numbers of expected background events, $b$. The smoothest curve is the coverage in the high statistics (normal) limit.}
\label{fig:NM2}
\end{figure}

\begin{figure}
\includegraphics[width=75mm]{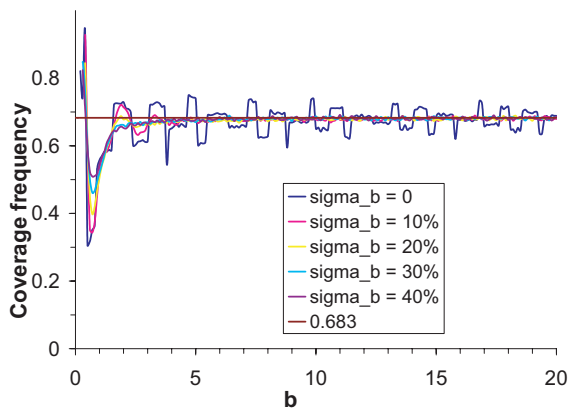}
\caption{Coverage frequency as a function of mean background $b$ for $B=0$, $f=1$, $\sigma_f=0$, $\Delta=1/2$. There are several curves corresponding to different values of $\sigma_b$, becoming smoother as $\sigma_b$ increases. The horizontal line is at 68\%.}
\label{fig:NM3}
\end{figure}

\begin{figure}
\includegraphics[width=60mm,angle=270]{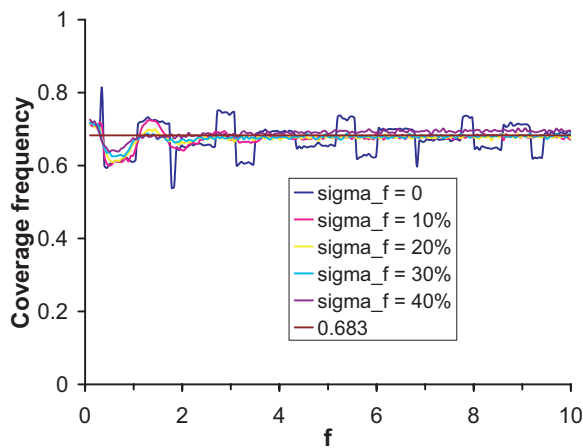}
\caption{Dependence of coverage on scale factor $f$ and $\sigma_f$ for $B=1$, $b=2$, $\sigma_b=0$, $\Delta=1/2$. There are several curves corresponding to different values of $\sigma_f$, becoming smoother as $\sigma_f$ increases. The horizontal line is at 68\%.}
\label{fig:NM4}
\end{figure}

\begin{figure}
\includegraphics[width=60mm,angle=270]{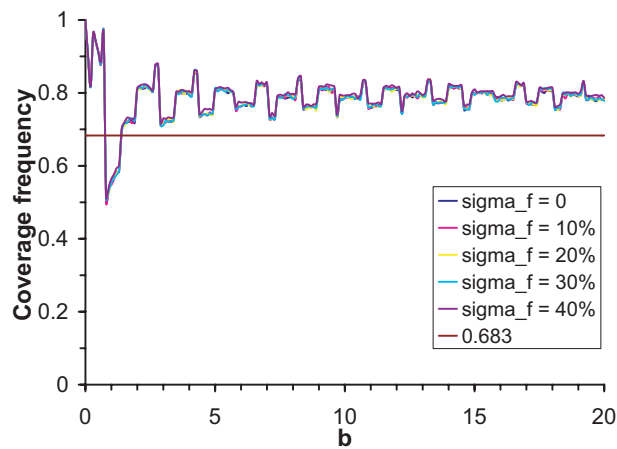}
\caption{Coverage as a function of expected background for $\Delta=0.8$, $B=0$, $f=1$, $\sigma_b=0$.  There are several curves corresponding to different values of $\sigma_b$, becoming smoother as $\sigma_b$ increases. The horizontal line is at 68\%.}
\label{fig:NM5}
\end{figure}

The question, of course, is: Does it work? To answer this, we need to investigate the frequency property of the algorithm. For large statistics (i.e., the normal limit) we know it works --- for $\Delta=1/2$ this method produces a 68\% confidence interval for $B$. We expect that it will fail in the extreme small statistics limit, and the question becomes a quantitative one of how far it can be pushed into the low statistics regime. We answer this with Figs.~\ref{fig:NM1}--\ref{fig:NM5}.

Figure \ref{fig:NM1} shows the dependence of the coverage of this algorithm on the value of $\Delta$, for several values of $B$ and an expected background of 1/2 event. The branching fraction scale is adjusted so that $B$ may be interpreted as the mean number of signal events. It may be seen that $\Delta=1/2$ gives coverage reasonably close to 68\% for $B\ge2$. Figure~\ref{fig:NM2} shows the coverage for $B=0$, for several backgrounds. Even at zero branching fraction, the $\Delta=1/2$ coverage is fairly close to 68\% for expected backgrounds $b\ge2$. Note that extending this to intervals with higher confidence may result in different conclusions.

It may be remarked that uncertainties in the background and/or scale factor help to obtain the desired coverage (Figs.~\ref{fig:NM3} and~\ref{fig:NM4}). This is because they smooth out the effect of the discreteness of the Poisson sampling space.

One issue is when the coverage is deemed to be ``good enough''. It might be suggested that if the coverage is known to be within some amount, say 5\% of 68\%, that this is good enough for anything we are going to use those numbers for. However, one could also decide to take a ``conservative'' approach, and insist that the coverage be at least at the quoted level. One way to accomplish this is to shift the value of $\Delta$. Fig.~\ref{fig:NM5} shows the coverage as a function of expected background (in the worst-case of zero signal branching fraction and $\sigma_b=0$) for a value of $\Delta=0.8$. We see that at least 68\% coverage is guaranteed as long as the mean background is greater than 1.4.

We'll conclude this discussion with a few summary remarks: First, it is a good idea to always quote $n, \hat b\pm\sigma_b$, and $\hat f\pm \sigma_f$. Second, any approach used should be justified with a computation of the coverage. The likelihood analysis studied here works pretty well even down to rather low statistics for 68\% confidence intervals. It should be kept in mind however that ``good enough'' for 68\% intervals does not imply good enough for other purposes, such as tests of significance. Finally, if $\sigma_b\approx b$ or $\sigma_f\approx f$ this is outside the regime studied here; the normal assumption is likely invalid in this case.

\subsubsection{Interpretation Intervals}

In the interpretation stage, Bayesian intervals may be given, as deemed useful to the consumer. In BaBar practice, this is typically done when someone wants to give an upper limit, and is usually implemented with the assumption of a uniform prior in the parameter of interest. BaBar recognizes the issues surrounding the choice of prior. The recommendation is to consider it carefully, and to make checks on how sensitive the result is to the choice. Even this recommendation is not routinely adopted however.

\subsection{Significance}

The ``significance'' of an observation (e.g., of the presence of a signal for some process) is defined as the probability of the observed deviation (or larger) from the null (no signal) model, under the null hypothesis. The recommended procedure in BaBar is to compute this probability according to the frequentist methodology. It may be noted that knowing the 68\% confidence interval does not always provide much insight into the significance. The tails of the null sampling distribution may be non-normal. A separate analysis is generally required, in which the tails are appropriately modelled.

No recommendation is tendered for when to label a result as ``significant''. We struggled with possible algorithms, but eventually gave up, because such a label implies an interpretation. No uniform prescription seems to make sense; judgement is involved. For example, deciding that the observation of a bizarre new particle is significant may involve a different standard than the claim that an expected decay mode of an established particle is significant. It isn't really our primary role as experimenters; it is up to the reader ultimately to decide what they wish to believe. This is perhaps the least-accepted of the Statistics Working Group's points in BaBar: people insist on making qualitative statements, e.g., ``observation of'', ``evidence for'', ``discovery of'', ``not significant'', ``consistent with''. A code exists in which ``observation of'' becomes quantified as $>4\sigma$ significance, and ``evidence for'' means $>3\sigma$.

This preoccupation with qualitative interpretive terminology is pervasive beyond BaBar. For example, the following excerpt appeared in Physics Today~\cite{bib:PT}, (italics mine, references deleted):

\vskip.1cm

\hskip0.3cm\vbox{\hsize=7cm\noindent
``In March, back-to-back papers in Physical Review Letters reported the measurement of CP symmetry violation in the decay of neutral B mesons by groups in Japan and California. Now the word ``{\it measurement}'' has been replaced by ``{\it observation}'' in the titles of two new back-to-back reports by these same groups in the 27 August Physical Review Letters. That is to say, with a lot more data and improved event reconstruction, the BaBar collaboration at SLAC and the Belle collaboration at KEK in Japan have at last produced the {\it first compelling evidence} of CP violation in any system other than the neutral K mesons.''
}

\vskip.2cm

For another example, some people think a measurement should not be called a ``measurement'' unless the result is significantly different from zero.
An editor at a prominent journal has suggested that ``{\it bounds on}'' might be more appropriate than ``{\it measurement}'' in reference to a CP asymmetry angle which was observed as consistent with zero. This can lead to amusing ironies:
Finding $\sin2\beta = 0.00 \pm 0.01$
would be an exciting contradiction with the standard model. But it isn't a ``measurement''?

A further issue that arises is that many people mix the question of significance with the choice of interval (i.e., one-sided vs two-sided). 
This has a drawback, because basing how one quotes the interval based on the result of the measurement can introduce a bias. The algorithm of Feldman and Cousins~\cite{bib:FeldmanCousins} is designed to address this. However, this methodology is not adopted in BaBar because of the constraint on the physical region, as discussed earlier. Instead, our recommendation is to always give a two-sided interval (if otherwise appropriate), independent of the significance. The significance is quoted separately. Quoting a one-sided interval may optionally also be done, and is usually regarded as part of the interpretation (hence a Bayesian approach is suggested). This recommendation is typically followed in BaBar, but there have been exceptions.

Another issue that arises in the quoting of significance has to do with the tradition of quoting significance as $n\sigma$. Unfortunately, this is used to mean different things: Sometimes it actually means $n$ standard deviations. But sometimes it means the probability content of an $n\sigma$ fluctuation for a normal distribution. We recommend to quote directly the probability if the sampling distirbution is not normal. However, this has met with very limited implementation.

\subsection{Systematic Uncertainties}

BaBar makes many checks in a typical analysis. For the purpose of defining systematic uncertainties, we divide these into two broad categories:

\begin{enumerate}
 \item ``Blind checks'': This is a test for mistakes. No correction to the data is anticipated. If the test passes, then there is no contribution to the systematic error. An example of such a check is dividing the data into two chronological subsets and comparing the results.
 \item ``Educated checks'': This is a measurement of biases or corrections, and may affect the quoted result. It involves a contribution to the systematic error. An example is the model dependence of the efficiency calculation.
\end{enumerate}

It is recommended that the systematic uncertainty be quoted separately from the statistical uncertainty. The sources of systematic uncertainty should be described, and may contain statistical components, for example due to limited Monte Carlo statistics in the efficiency evaluation.

\begin{figure}
\includegraphics[width=75mm]{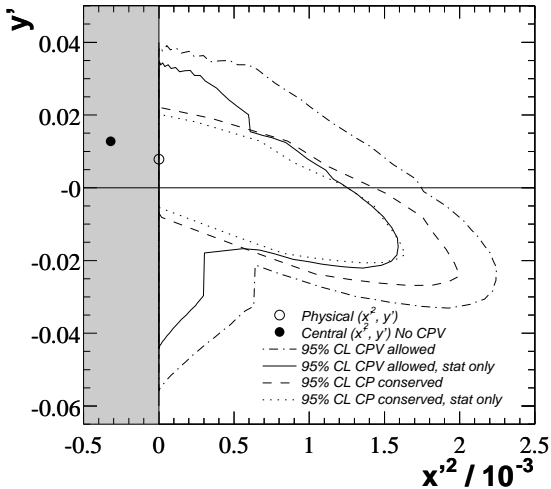}
\caption{Incorporating systematic uncertainties in the confidence contour for $D$ mixing. The filled dot is the location of the best fit point, $(\hat x^{\prime 2}, \hat y^\prime)$, the open circle the best fit point in the physical region. The solid contour (dotted if restricting to $CP$ conserving models) shows the 95\% confidence contour according to statistical errors only. The dot-dash contour (dash for $CP$ conserving models) shows how this contour becomes scaled on incorporating systematic uncertainties.}
\label{fig:DmixSyst}
\end{figure}

We return to our earlier example (Sec.~\ref{sec:Dmix}) of $D$ mixing for an example of the treatment of systematic uncertainties. The goal here is to produce a two-dimensional confidence contour in the parameter space which incorporates the systematic uncertainites. In this case, the statistical uncertainties are large, and we are willing to accept an approximation in order to keep the procedure simple. Thus, it is decided to use a method which takes the statistics-only contour and scales it uniformly along rays from the best fit value. The scaling factor is $\sqrt{1+\sum m_i^2}$, where $m_i$ is an estimate of systematic uncertainty $i$ in units of the statistical uncertainty. This estimate is obtained by determining the effect of the systematic uncertainty on $\hat x^{\prime 2}, \hat y^\prime$ (the position of the best fit). Figure~\ref{fig:DmixSyst} shows the result of this procedure. This method is conservative (or lazy) in the sense that scaling for a given systematic in one (worst case) direction is applied uniformly in all directions. On the other hand, by evaluating the error at the best fit position, a linear approximation is being made.

\subsection{Goodness of Fit}

There appears to be no perfect general goodness-of-fit test.  Given a dataset generated under the null hypothesis, one can usually find a test which rejects the null hypothesis (and this may be taken as a warning that choosing the test after you see the data is dangerous). Given a dataset generated under an alternative hypothesis, one can usually find a test for which the null passes. It seems advisable to think about what one wants to test for in choosing the test.

\begin{figure}
\includegraphics[width=85mm]{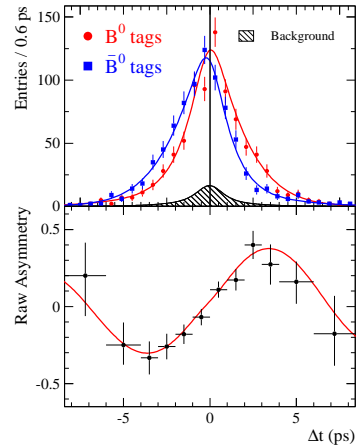}
\caption{Measurement of $CP$ violation (BaBar)~\cite{bib:sin2beta}. The upper plot shows the measurement (points) of the time distributions for $B^0$ and $\bar B^0$ decays to selected $CP$ eigenstates. The curves show the result of a maximum likelihood fit to the data. The lower plot shows the time-dependent asymmetry between the $B^0$ and $\bar B^0$ decays, again with the fitted curve overlaid. The asymmetry would be zero in the absence of $CP$ violation.}
\label{fig:CPviol}
\end{figure}

For example, Fig.~\ref{fig:CPviol} shows data used in a measurement of $CP$ violation by BaBar. A likelihood ratio (or a chi-square) test of the time distribution may be a good test for the lifetime fit to the data, but it may have little sensitivity to testing the goodness-of-fit of the $CP$ asymmetry, which is a low-fequency question. 

So far, BaBar generally uses likelihood ratio tests or chi-square tests if appropriate. The Kolmogorov-Smirnov test is also used. If a test statistic such as the likelihood ratio is used, then a Monte Carlo evaluation of the distribution of the statistic is recommended, rather than assuming an asymptotic property.

\subsection{Consistency of Analyses}

BaBar has encountered several times the question of whether a new analysis is consistent with an old analysis.
Often, the new analysis is a combination of additional data plus changed (improved) analysis of original data. 
The stickiest issue is handling the correlation in testing for consistency in the overlapping data.
People sometimes have difficulty understanding that statistical differences can arise even comparing results based on the same events, so we expound on this.

Given a sampling $\hat\theta_1,\hat\theta_2$ from a bivariate normal distribution $N(\theta,\sigma_1,\sigma_2,\rho)$, with $\vev{\hat\theta_1}=\vev{\hat\theta_2}=\theta$, the difference $\Delta\theta\equiv\hat\theta_2-\hat\theta_1$ is $N(0,\sigma)$-distributed with $\sigma^2 = \sigma_1^2 +\sigma_2^2 -2\rho\sigma_1\sigma_2$. 
If the correlation is unknown, all we can say is that the variance of the difference is in the range $(\sigma_1-\sigma_2)^2\ldots(\sigma_1+\sigma_2)^2$. If we at least believe $\rho\ge0$ then the maximum variance of the difference is $\sigma_1^2+\sigma_2^2$.

Suppose we measure a neutrino mass, $m$, in a sample of $n=10$ independent events. The measurements are $x_i, i=1,\ldots,10$. 
Assume the sampling distribution for $x_i$ is $N(m,\sigma_i)$.

We may form {\it unbiased} estimator, $\hat m_1$, for $m$:
\begin{equation}
 \textstyle{\hat m_1 = {1\over n}\sum_{i=1}^n x_i \pm \sqrt{{1\over n^2}\sum_{i=1}^n \sigma_i^2}}.
\end{equation}
The result (from a Monte Carlo simulation) is $\hat m_1 = 0.058 \pm 0.039$.

Then we notice that we have some further information which might be useful: we know the experimental resolutions, $\sigma_i$ for each
measurement. We form another  {\it unbiased} estimator, $\hat m_2$, for $m$: 
\begin{equation}
 \textstyle{\hat m_2 = \sum_{i=1}^n {x_i\over \sigma_i^2}/\sum_{i=1}^n {1\over \sigma_i^2} \pm 1/\sqrt{\sum_{i=1}^n {1\over\sigma_i^2}}}.
\end{equation}
The result (from the same simulation, i.e., from the {\it same events}) is $\hat m_1 = 0.000 \pm 0.016$.

The results are certainly correlated, so the question of consistency arises (we know the error on the difference is between 0.023 and 0.055). In this example, the difference between the results is
$0.058 \pm 0.036$, where the $0.036$ error includes the correlation ($\rho=0.41$).

Art Snyder has developed an approximate formula for evaluating the correlation in a comparison of maximum likelihood analyses.
Suppose we perform two maximum likelihood analysis, with event likelihoods ${\cal L}_1$, ${\cal L}_2$, on the same set of $N$ events [n.b., we may use different information in each analysis]. The results are estimators $\hat\theta_1$, $\hat\theta_2$ for parameter $\theta$ (restricting to the one-dimensional case for simplicity).
The correlation coefficient $\rho$ may be estimated according to:

\begin{equation}
\rho \approx {\sum_{i=1}^N R_i {d\ln{\cal L}_{1i} \over d\theta}\vert_{\theta=\hat\theta_1} {d\ln{\cal L}_{2i} \over d\theta}\vert_{\theta=\hat\theta_2} \over \sqrt{\left(\sum_{i=1}^N {d^2\ln{\cal L}_{1i} \over d\theta^2}\vert_{\theta=\theta_0}\right)
\left(\sum_{i=1}^N {d^2\ln{\cal L}_{2i} \over d\theta^2}\vert_{\theta=\theta_0}\right)}},
\label{eqn:consistency}
\end{equation}
where ($\theta_0$ is an expansion reference point):
\begin{eqnarray}
R_i=&&\hskip-.4cm\left[1-(\hat\theta_1-\theta_0) {d^2\ln{\cal L}_{1i} \over d\theta^2}\vert_{\theta=\theta_0}\Bigg/{d\ln{\cal L}_{1i} \over d\theta}\vert_{\theta=\theta_0} \right]\nonumber \\
 &&\hskip-.4cm\left[1-(\hat\theta_2 - \theta_0) {d^2\ln{\cal L}_{2i} \over d\theta^2}\vert_{\theta=\theta_0}\Bigg/{d\ln{\cal L}_{2i} \over d\theta}\vert_{\theta=\theta_0} \right].\nonumber
\end{eqnarray}
If $\theta_0\approx\hat\theta_1\approx\hat\theta_2$, then
\begin{equation}
\rho \approx \tilde\sigma_{\theta_1} \tilde\sigma_{\theta_2} \sum_{i=1}^N {d\ln{\cal L}_{1i} \over d\theta}\vert_{\theta=\theta_0} {d\ln{\cal L}_{2i} \over d\theta}\vert_{\theta=\theta_0},
\end{equation}
where
$\tilde\sigma_{\theta_k}^2 \equiv 1/\sum_{i=1}^N\left({d{\cal L}_{ki} \over d\theta}\vert_{\theta=\theta_0}\right)^2$.

Let us look at a real example of the consistency question in a BaBar analysis, the measurement of the $CP$-violation parameter $\sin2\beta$. In August 2001, we published a result based on a dataset of $32\times 10^{6}\, B\bar B$ pairs~\cite{bib:sin2betaA}:
\begin{equation}
\sin2\beta = 0.59\pm0.14({\rm stat})\pm 0.05({\rm syst})
\end{equation}
An updated result was produced in March 2002, based on $62\times 10^{6}\, B\bar B$ pairs~\cite{bib:sin2betaB}:
\begin{equation}
 \sin2\beta = 0.75\pm 0.09({\rm stat})\pm 0.04({\rm syst})
\end{equation}
The second result includes the earlier data, re-reconstructed. The analysis is not simply counting events; it involves multivariate maximum likelihood fits, reprocessing changes, and relative likelihoods for an event to be signal or background, for example. 
The question is, are the two results statistically consistent?

If these were independent data sets, a difference of $0.16\pm 0.17$ would not be a worry. The issue is the correlation.
A specialized analysis deriving from Eqn.~\ref{eqn:consistency} is performed on the events in common between the two analyses.
A correlation of $\rho=0.87$ is deduced, yielding a difference of $\sim 2.2\sigma$. This corresponds to a probability of 3\%, which is small enough that we noticed, and looked hard for possible systematic problems, but not so small to be alarming, especially in  an experiment with many such tests being made.

There has been some impression that BaBar may be seeing more diffences between old vs updated results than people are used to, and the question arises whether BaBar is making mistakes. The answer to this seems to be, first of all, based on studies such as the above, there is no compelling statistical evidence to support the contention that mistakes are being made. There should be differences, purely due to statistical fluctuations, among results, and BaBar sees nothing clearly beyond what might be expected from statistics. The second part of the answer is a speculation to why the impression may exist. BaBar is different from most other experiments in that it makes extensive use of the blind methodology. There is little opportunity to react to observed differences with further changes in analysis. Without using the blind methodology, there is the potential for bias, tending towards making results agree with earlier results better than they should.

\section{Reflections}

It is my observation that statistical sophistication in particle physics (not specific to BaBar) has grown significantly, not so much in the choice of methods, which are often long-established, but in the understanding attached to them. People now understand that there is a choice of approach between Bayesian and frequency statistics, though there is yet no uniform agreement on adoption. There is also considerable awareness on the issue of biases in analyses, for example, BaBar now relies heavily on blind methodology.

BaBar adopts frequency statistics for describing results, and much attention is devoted to Monte Carlo validation and verification of coverage. The use of the Bayesian approach in high energy physics, including BaBar, is still not mature: There is no established methodology for choosing the prior distribution, other than to default on a uniform prior. The justification for this is basically that it usually doesn't matter very much. There are, however, even issues still in frequency statistics. Controversies involve such notions as restricting to the ``physical region'', or that the presence of backgrounds should ``always'' lead to higher upper limits. Both of these notions are not a concern in the BaBar recommendations.

BaBar is attempting to provide a coherent, documented approach to its use of statistics in its results. This is very much a work in progress.

%
%

\begin{acknowledgments}
I would like to thank Louis Lyons for organizing an informative and stimulating conference.
I am grateful to my collaborators on BaBar for many interesting discussions of statistical issues.
Work supported in part by Department of Energy grant DE-FG03-92ER40701 and contract DE-AC03-76SF00515.
\end{acknowledgments}

\end{document}